\documentclass[doublespace,11pt]{article}
\usepackage{graphicx}
\usepackage{setspace}
\usepackage{mathptm}
\usepackage{amsmath, amsthm}
\usepackage{subfigure}
\usepackage{epsfig}
\usepackage{url}

\begin{document}
\sloppy
\begin{spacing}{1}

\begin{titlepage}
\hspace{0.08in}
\begin{minipage}{\textwidth}
\begin{center}
\vspace*{3cm}
\begin{tabular}{c c c}
\hline
 & & \\
 & {\Huge The University of Algarve} & \\
 & & \\
 & {\Huge Informatics Laboratory} & \\
 & & \\
\hline
\end{tabular}\\
\vspace*{2cm}
{\Large
UALG-ILAB\\
Technical Report No. 200604 \\
June, 2006\\
}
\vspace*{3cm}
{\bf Breaking barriers for people with voice disabilities:\\
Combining virtual keyboards with speech synthesizers, and VoIP applications}\\
\addvspace{0.5in}
{\bf Paulo A. Condado} and {\bf Fernando G. Lobo}\\
\vspace*{-0.1in}
\vspace*{4cm}
Department of Electronics and Informatics Engineering\\
Faculty of Science and Technology \\
University of Algarve \\
Campus de Gambelas\\
8000-117 Faro, Portugal\\
URL: {\verb http://www.ilab.ualg.pt }\\
Phone: (+351) 289-800900\\
Fax: (+351) +351 289 800 002 \\
\end{center}
\end{minipage}
\end{titlepage}

\title{\bf Breaking barriers for people with voice disabilities:
Combining virtual keyboards with speech synthesizers, and VoIP applications}
\author{    
            {\bf Paulo A. Condado}\\
            \small UAlg Informatics Lab\\
            \small DEEI-FCT, University of Algarve\\
            \small Campus de Gambelas\\
            \small 8000-117 Faro, Portugal\\
            \small pcondado@ualg.pt
\and
          {\bf Fernando G. Lobo}
\footnote{Also a member of IMAR - Centro de Modela\c{c}\~{a}o Ecol\'{o}gica.}\\
            \small UAlg Informatics Lab\\
            \small DEEI-FCT, University of Algarve\\
            \small Campus de Gambelas\\
            \small 8000-117 Faro, Portugal\\
            \small flobo@ualg.pt
}
\date{}
\maketitle

\begin{abstract}
Text-to-speech technology has been broadly used to help people with voice 
disabilities to overcome their difficulties. With text-to-speech, a person
types at a keyboard, the text is synthesized, and the sound comes out 
through the computer speakers.

In recent years, Voice over IP (VoIP) applications have become very popular
and have been used by people worldwide. These applications allow people to 
talk for free over the Internet and also to make traditional calls through the 
Public-Switched Telephone Network (PSTN) at a small fraction of the cost 
offered by traditional phone companies.

We have created a system, called {\em EasyVoice}, which integrates 
speech synthesizers with VoIP applications. The result allows a person
with motor impairments and voice disabilities to talk with another person
located anywhere in the world. The benefits in this case are much stronger 
than the ones obtained by non-disabled people using VoIP applications. 
People with motor impairments sometimes can hardly use a regular or 
mobile phone. Thus, the advantage is not only the reduction in cost, but 
more important, the ability to talk at all.
\end{abstract}

\section{Introduction}
\label{sec:intro}
Communication is one of the most important factors for the development of a 
Human Being. Every person needs to communicate with the surrounding world 
because each one of us needs to be able to exchange ideas and experiences. 
The exchange of knowledge is the basis of our society, but
unfortunately, many people with physical disabilities are unable to 
communicate easily with theirs friends, teachers, colleagues, and family
members.

Information and communication technologies are becoming an integral
part of our society. These new technologies are changing all sectors of our 
society, and have been used to improve the quality of life of people with 
physical disabilities. With the advances in computing power, new applications 
have been developed for helping these people. Voice synthesizers 
are a good example of these applications. Voice synthesis quality and the 
associated computational resources have advanced considerably in the last 
years \cite{Condado:2003, Dutoit:2003}, and have been a key factor 
for a better integration of people with voice disabilities in society.



People without disabilities can easily make phone calls. However, 
when we are talking of disabled people, a phone call can be something almost 
inaccessible. Many times, people with voice disabilities are dependent on a 
friend, a family member, or a colleague, to communicate with someone that is 
in another location.

This paper presents the EasyVoice system. 
EasyVoice is a Text-to-Speech (TTS) interface that provides an easy way 
for typing the desired messages to be synthesized. The application, when 
properly used with an external driver and a VoIP application, allows 
people with disabilities to call anywhere.

The paper is organized as follows. The next section present background
material which motivated the present work: Text-to-speech technology, 
virtual keyboards, and voice over IP applications. 
Then, section~\ref{sec:easyvoice} presents the EasyVoice
system, an architecture that combines existing technologies to create
a novel functionality for people with voice disabilities. The paper ends
with an outline of future work in section~\ref{sec:futurework}, and with 
summary and conclusions in section~\ref{sec:conclusions}.

\section{Background}
\label{sec:background}
This section describes background material which constitute the foundations
of the present work. We start by briefly describing our past experience with
text-to-speech systems. Then, we describe previous research on user 
interfaces specifically designed to provide people with motor impairments 
the ability of writing at a reasonable speed in a computer.
Finally, we mention briefly mention Voice over IP applications.

\subsection{Text-to-Speech technology}
\label{sec:tts}
Our first experience with text-to-speech occurred a few years ago.
At the time, a student with cerebral palsy needed to give 
a presentation about his final undergraduate project at the University 
of Algarve. These kind presentations are public, and it is common to have many 
people in the audience, including colleagues, friends, and family members. 
Since not many people in the audience could understand the student's voice,
a colleague of ours, Hamid Shahbazkia, thought about using speech synthesis 
for the presentation. This way, the student could type what he wanted to say 
and the computer could synthesize the text and deliver the corresponding audio 
to the audience \cite{Condado:2003}. 

The speech synthesizer that was developed at the time was based on the
Mbrola project \cite{Dutoit:96}, a research work developed by Dutoit et 
al.~whose goal was to provide high quality speech synthesizers free of use 
for non-commercial purposes. We started by studying the Mbrola TTS and tried
to adapt it to the Portuguese language. The system can be adapted by creating
rules which, in their simplest form, specify how particular sequences of 
letters are synthesized.
In our study we found that about one hundred rules could be successfully 
applied to achieve a good TTS for the Portuguese language.
But more rules could be easily added to improve the overall quality of the system.


A transformation computer program was implemented. It takes as 
input a complete Portuguese text, transforms it and sends it to the Mbrola 
system. As an addition to the imposed rules, we also utilized a random pitch making 
the output sound more human, introducing in this way an unexpected factor, since 
the vocal human sounds are the result of our lung pressure, glottis tension, and 
vocal and nasal tracts, among many others. The software initially loads all the 
imposed rules, reads the input text, transforms the text to lower case (since 
the Mbrola treats upper and lower case letters in a different way) and searches 
for the groups of letters, or words, where a rule can be applied. Since the output 
of each transformation can have another group of letters where another rule 
can be applied, a loop in the transformation procedure is done. Also, we needed 
to be careful with rules that generate infinite transformations, and to impose a 
looping limit.

The rules are based on word context and phonetic duration. The word context is 
done by checking the surrounding character or characters of the candidate rule. 
The rule definition file uses the symbol '*' to represent all vowels, '+' to
represent all consonants, and '-' to represent a space. As an example, the rule:

\begin{verbatim}
que k@ 100 | -
\end{verbatim}

Means that the groups of letters {\tt 'que'} should be changed to {\tt 'k@'} with 
duration 100\% of the speed (not changing the current output speed), if followed 
by a space. In fact, the {\tt 'k@'} sounds much better than the {\tt 'que'}. 
As another example, the rule:

\begin{verbatim}
x z 60 a e o u | *
\end{verbatim}

Changes all the 'x' to 'z', with duration 60\% of the normal duration, if the x 
is followed by a vowel and if it is preceded by an 'a' or 'e' or 'o' or 'u'.

Once the prototype system was built, the student with cerebral palsy had the 
opportunity to practice with it. The system allowed him to either type and 
get the voice synthesis in real time, or use the system to synthesize a text 
file previously stored on the computer. The final presentation was a success, 
with his talk being delivered in a mixed way, sometimes using the voice 
synthesizer, and sometimes using his own voice. 

A person with voice disabilities usually also has motor disabilities.
That's often the case of people with cerebral palsy. In such cases, typing
at a regular keyboard can also be a difficult thing to do, and that is an
obstacle for a smooth utilization of a speech synthesizer. Because of that,
special purpose user interfaces need to be considered.

\subsection{Virtual keyboards}
\label{sec:virtualkeyboard}
People with motor disabilities need an easy and accessible interface for 
computer interaction. For example, many individuals cannot control
their hands with enough accuracy to use a regular computer keyboard,
and sometimes only have the ability to control a single touch button.

Virtual keyboards are a reasonable solution to solve 
some of these limitations. Most virtual keyboards have a set of features 
to accelerate the writing process. Keyboard group scanning function and word 
prediction systems are among the most common features to achieve that.

With a scanning system, a set of options is presented to the user on the
computer screen, and a visual cursor advances through the options, one at time,
at a specified time rate. The user responds by pressing a touch button whenever
the cursor is on top of the desired option.
Sometimes an option is just a container for more options, in which case the 
options are organized in a hierarchical fashion. 
Each container option is often referred to as a {\em group option}.
For example, when a particular group option is selected, the scanning system 
immediately focuses on the sub-options of that group option, and again, 
advances the visual cursor through each of them \cite{Demasco:92,Condado:2004}.

Word prediction techniques are also used for accelerate the writing process.
Most mobile phones have this feature incorporated for allowing people to type
text messages faster. When the user types a letter, or a set of letters, the 
system shows a list of possible words that have those letters as a prefix. 
The system does that by searching in a built-in dictionary, which can also be 
updated with new words. 
This feature allows the user to select the desired word without having
to type every single letter of it, decreasing the necessary time for 
writing a message \cite{Boissiere:2003}.


We have used these techniques in the past to build a virtual keyboard for 
our Mbrola based TTS 
\cite{Condado:2004}. 
The resulting keyboard gives the opportunity for people with severe motor 
disabilities to communicate more easily and faster.

Nevertheless, it is not always the case that people with voice disabilities 
also have severe motor problems. Sometimes, people only have minor motor
problem (or even no motor problems at all). In such cases, they might 
not want to use the virtual keyboard for TTS, because 
typing directly on a physical keyboard might be faster in those cases.
It is with these ideas in mind that we developed the EasyVoice TTS interface.
EasyVoice has a built-in virtual keyboard, but it can be turned on or off
according to the user's preferences.

\subsection{Communication with Voice over IP}
\label{sec:voip}
Voice over IP (VoIP) applications have become very popular during the last few
years. VoIP has established a commercial niche after the mid-1990s \cite{Varshney:2002}. 
The first commercial VoIP product was produced in 1995 and it did not allow calls 
for regular phones. In subsequent years, however, the commercial VoIP products, 
as well as the quality of Internet connections, have improved considerably. 

VoIP is a useful technology for millions of people worldwide. With VoIP applications
such as Skype, MSN, Google Talk, or VoipBuster, people can call each other at a
very small fraction of the cost of a regular phone call. 
The cost reduction is useful on its own, but when we think about people with 
voice disabilities, there is not only the cost advantage, but more important, there
is the advantage of being able to communicate, something that might have
not been possible for those with severe motor problems. To many people, the simple
task of picking a phone and dialing a number, is something nearly impossible to 
achieve.

\section{The EasyVoice system}
\label{sec:easyvoice}
\begin{figure}
\center
\epsfig{file=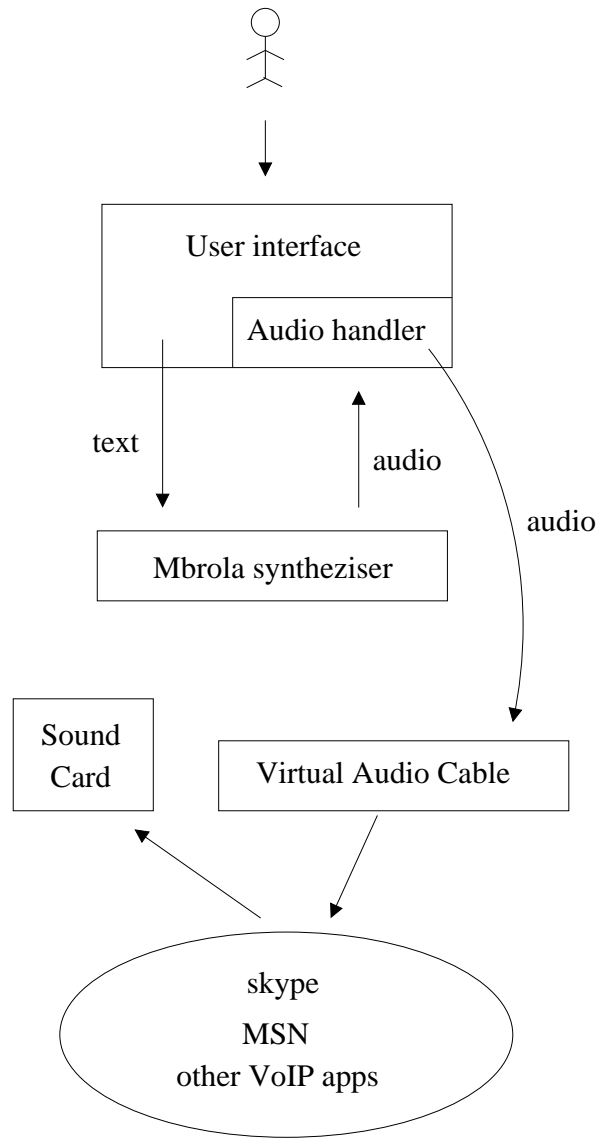,width=0.5\columnwidth}
\caption{The architecture of the EasyVoice system.}
\label{fig:scheme}
\end{figure}

We have reviewed various technologies in the previous section.
Having this in mind, we thought about integrating a speech synthesizer with VoIP
applications, and provide an appropriate user interface for people with motor 
impairments. We named the system {\em EasyVoice}, and its architecture is 
depicted schematically in Figure~\ref{fig:scheme}.

Figure~\ref{fig:scheme} shows how a set of independent applications can work 
together to do a specific objective. Each application has its own functionality
and can work independently of each other. However, they can also work
together, creating something that is more than the sum of its parts.
In our opinion this is a good example of systems integration and good engineering 
principles. Rather than building a whole new application from scratch, we
combined existing applications in a novel way.

The interaction between these different software systems is as follows:

\begin{enumerate}
 \item The EasyVoice TTS interface receives text written by a person.
 \item The system searches by abbreviation, replaces each abbreviation by the 
       full spelled word, transform the entire text in a string readable by the 
       Mbrola's phonetic Portuguese database, and sends the text to the Mbrola
       synthesizer.
 \item The Mbrola system converts the input text into an audio file.
 \item The EasyVoice TTS interface plays the audio file which is intercepted by 
       a virtual audio cable (VAC).
 \item VAC allows us to transfer audio (WAV) streams from one application to 
       another. This way, the audio file generated by Mbrola can be played by
       the EasyVoice interface and sent to a VoIP application in a transparent
       way.
\end{enumerate}

\subsection{The EasyVoice text-to-speech user\\interface}
\label{sec:easyvoice-tts}
We have developed EasyVoice, a TTS interface that uses the Mbrola 
synthesizer \cite{Dutoit:96}. In principle, other synthesizers could be used as well.
With EasyVoice, users can choose if they want to write directly on a text box area 
or to open a virtual keyboard. By making the utilization of the virtual keyboard an 
optional feature, rather than a mandatory one, the user interface becomes more
flexible and useful for a wider audience. Those who have voice disabilities
but do not have severe motor problems, can choose to type in the normal way. 
The virtual keyboard, on the other hand, can be selected for helping users with 
motor disabilities, or to allow the application to be used in touch 
screen machines.

The EasyVoice user interface has four main features:

\begin{itemize}
\item an archive of recent messages.
\item a word prediction system.
\item an abbreviation system.
\item an optional virtual keyboard.
\end{itemize}

\begin{figure}
\center
\epsfig{file=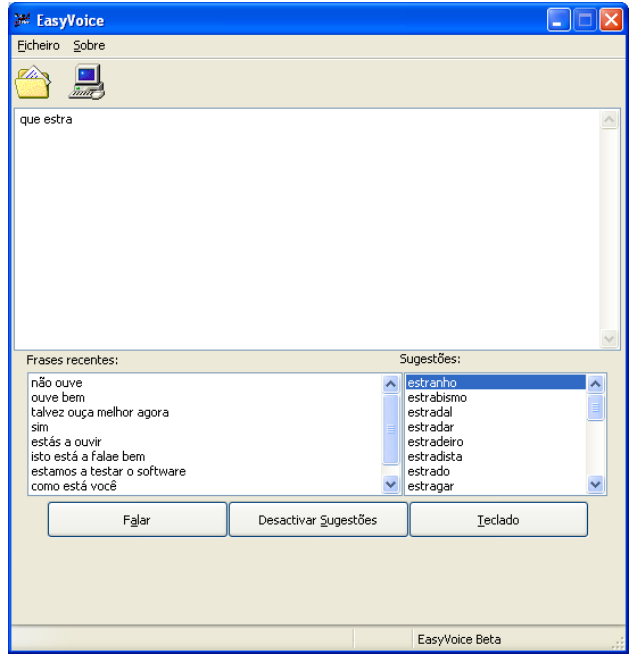,width=0.5\columnwidth}
\caption{The EasyVoice user interface.}
\label{fig:easyvoice}
\end{figure}

Figure~\ref{fig:easyvoice} shows a screenshot of the EasyVoice user interface.
There is a text panel on the top part where the user can type. Below the text
panel there are two list boxes. The one on the left part is a list of recent 
messages, and on the right part is a list of possible words given by the
word prediction system. 

The archive of recent messages is a very important feature because during
a conversation it is many times necessary to repeat some words or phrases,
because the other person at the end of the line might have not heard the 
sentence well enough. With the archive in hand, the user does not need to
retype the message and can simply pick it from the list again.

The word prediction system that we have implemented is based on a simple
algorithm. We have a built-in Portuguese dictionary with 32334 words. 
The prediction system searches the dictionary for those words that have 
as a prefix the sequence of letters typed by the user so far. 
Alternative prediction systems could have also been used instead.
For example, rather than having a very large built-in dictionary,
with a large fraction of the vocabulary never being utilized, we
could have a small dictionary with just 100-200 highly frequent words from
the language. Then, the user could insert new words in the dictionary
through time. This way, the word prediction system would be tailored to 
the user \cite{Boissiere:2003}. Notice that different people have different 
writing and talking styles, and may use some words very frequently while
other users may not use them at all. A solution like that would increase 
the efficiency of the prediction system considerably, and it is something 
that we have in mind for future work.

\begin{figure}
\center
\epsfig{file=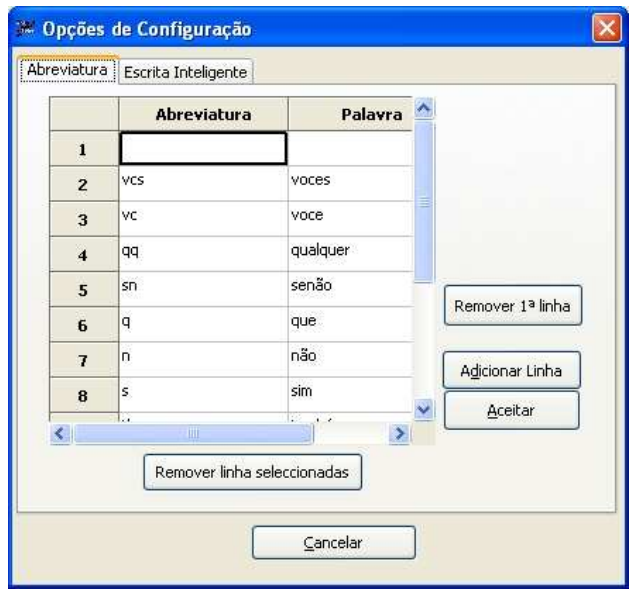,width=0.5\columnwidth}
\caption{The abbreviation system of EasyVoice.}
\label{fig:abreviations}
\end{figure}

Another important feature is the abbreviation system. Every person uses 
abbreviations in their daily life, and it is something that is very popular
in instant messaging software, especially among young people. For example,
it is common for Portuguese people to type ``{\tt qq}'' as an abbreviation
for the other ``{\tt qualquer}''. Likewise, in English it is common for
people to use ``{\tt btw}'' as an abbreviation of ``{\tt by the way}''.
Within EasyVoice, the user can define his own abbreviations 
(see Figure~\ref{fig:abreviations}). The system
will automatically replace each abbreviation by the full spelled word,
before sending it to the speech synthesizer.

\begin{figure}
\center
\epsfig{file=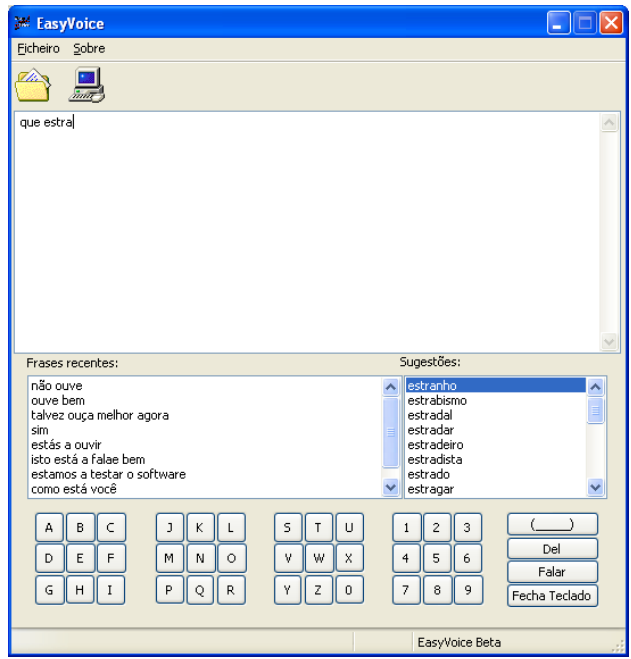,width=0.5\columnwidth}
\caption{The EasyVoice user interface with the virtual keyboard option turned on.}
\label{fig:easyvoice-with-keyboard}
\end{figure}

The last feature of the EasyVoice user interface is a a virtual keyboard.
When this feature is turned on, a keyboard pops down and the user interface
becomes like the one shown in Figure~\ref{fig:easyvoice-with-keyboard}.

\begin{figure}
\center
\epsfig{file=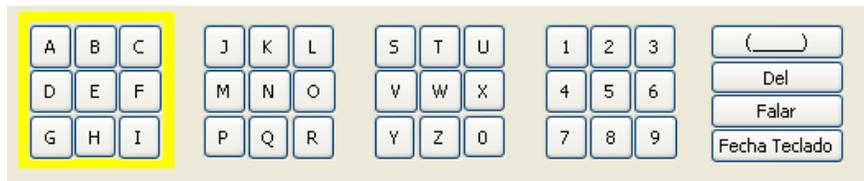,width=0.7\columnwidth} 
\caption{An example of the scanning system. There are 5 groups of keys, and the cursor
advances through each group at a specified time rate.}
\label{fig:group}
\end{figure}

\begin{figure}
\center
\epsfig{file=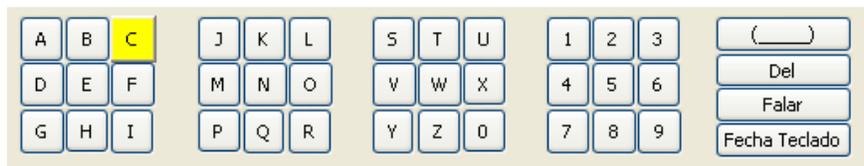,width=0.7\columnwidth} 
\caption{In this example, the first group was selected, and now, the scanning 
system becomes focuses on the keys of that group, with the cursor advancing 
through each of the 9 letters (A,B,C,D,E,F,G,H,I), one at a time.}
\label{fig:key}
\end{figure}

The keyboard does not have a qwerty format because we believe that an alphabetic 
order format is more accessible for people with disabilities. The keyboard has
a group scanning mechanism like the one described in section~\ref{sec:virtualkeyboard},
and which we describe again schematically in Figures~\ref{fig:group} and~\ref{fig:key}.

With the virtual keyboard incorporated, the EasyVoice user interface can be used 
by people with motor disabilities, and can also be used at Internet 
kiosks by anyone.

\subsection{Voice over IP communications and EasyVoice}
\label{sec:voip-easyvoice}
The most important concept behind this work is the idea of communication.
The idea that people with voice disabilities are not limited to a room, 
and that they can use voice synthesizers to communicate with entire world. 
This idea integrates many concepts such as text-to-speech synthesizers, the Internet, 
VoIP, and user interfaces. 

When we are talking of people with disabilities, a phone call can be an inaccessible 
thing. Fortunately, voice synthesis systems and VoIP programs can be used 
together for overcoming such limitations. At present time, one of most popular VoIP 
applications is Skype (\url{http://www.skype.com}), which provides VoIP and 
instant messaging services, and works behind firewalls and Network Address 
Translators \cite{Guha:2006}. 
The integration of TTS synthesizers, like ours, with Skype, or similar programs, 
can change the life of thousands of people because they can now speak with anyone,
anywhere.

The EasyVoice system uses a software called a Virtual Audio Cable (VAC) 
(see Figure~\ref{fig:scheme}) which was developed by Eugene Muzychenko
(\url{http://software.muzychenko.net/eng/}).
VAC is a Microsoft Windows multimedia driver allowing us to transfer 
audio (wave) streams from one application to another. It creates a set 
of ``Virtual Cables'' each of them consisting of a pair of the waveform input/output 
devices. Any application can send audio stream to an output side of a cable, and any 
other application can receive this stream from an input side. All transfers are 
made digitally, and there is no sound quality loss. 

The EasyVoice user interface, when properly configured to use a driver such as VAC,
can directly send the output generated from the speech synthesizer to a VoIP Software.

Communication for people with disabilities are becoming easier with new technologies. 
Sometimes, existing technologies must be observed from another point of 
view to evaluate their potential to do other functionalities. Some 
technologies, such as computers themselves, were originally created for helping 
human beings in their work and leisure activities. But these technologies can also
be used to help people with disabilities to overcome their difficulties. Of course, 
some specific interfaces need be adapted to provide a pleasant interaction.   

In our particular case, the interaction between voice synthesizers and other 
applications can open a new kind of perspectives for helping people with voice 
disabilities. A voice system always gives autonomy of speech, but sometimes that 
autonomy is limited due to physical barriers. These barriers are 
breaking down when a person with voice disabilities can make a phone call 
using voice systems. A simple phone call is something that people without voice 
disabilities take for granted, but that represents a dream to many people with voice 
disabilities. This dream is becoming a reality with EasyVoice. 
Thus, every person can communicate 
with friends, colleagues, and relatives, using an Internet connection, a 
VoIP software, and a voice system. When the user can use an Internet connection 
to keep in touch with other people a set of new perspectives appear and a window 
is open for a new world of communication, learning, and socialization.

Technology, learning, and socialization are very important factors because 
technology can help people to have a better education, and people with a good 
education are better integrated in society. 

\subsection{EasyVoice and learning}
Nowadays, communication between students is very 
important in classrooms because they need to exchange experiences and ideas. 
Constructionist theories of education, advocated by Seymour Papert, argues 
that children learn by exploring their world. Children are constructors of their 
own knowledge, and they learn with their own mistakes \cite{Papert:93,Papert:96}. 
Communication is an important component on that theory, because students are no 
more ``passive'' receptors of knowledge. They are ``knowledge makers'' and 
information needs to be shared between all students and their teachers.

Technology provides a new kind of education. An education based on Papert's 
theories where every student can use computers to investigate any given topic.
Collaborative works are a reality on that kind of learning. A student 
needs to communicate with his/her colleagues even at home. Phone calls are a 
fantastic way of communication, however a student with voice disabilities is 
very limited to do that. EasyVoice can break down these limitations and offer 
the possibility of a full communication.

Distance learning can be an option too for people with disabilities. Many times, 
people with severe physical disabilities cannot go to school because of their 
physical barriers. However, they can use video-conference technology to 
participate in lessons. EasyVoice and VoIP technology can help people with voice 
disabilities in distance learning. Distance is no more a barrier or an obstacle 
because disabled people can participate in ``regular'' lessons from their own 
homes.

When people with disabilities are well integrated in learning environments, 
they have a better chance of being well integrated in society. EasyVoice can 
break down communication barriers and give the same possibility to people with, 
and without, disabilities. The life quality of these people is improved considerably 
when communication becomes easier and faster. With EasyVoice TTS interface, they 
have a tool to communicate with their friends, family, teachers, bosses, and 
colleagues at anyplace. It is possible because the 
EasyVoice TTS interface can be used in small and portable computer systems, such 
as Toshiba Libretto \cite{Condado:2003}, that can be carried easily.

\subsection{EasyVoice and cerebral palsy}
\label{sec:cerebral-palsy}
Cerebral palsy is a condition that affects motor ability of a person. That motor 
disability is caused by a non-progressive lesion in the brain \cite{Miller:2004}. 
We have conducted experiments in collaboration with ``Associa\c{c}\~{a}o 
Portuguesa de Paralisia Cerebral'' (APPC), the Portuguese Cerebral Palsy 
Association. Tests conducted with a Voice Touch virtual keyboard have proved 
that people with disabilities, who cannot speak and have severe motor 
disorders, are able to communicate using adapted voice synthesizer systems. 
EasyVoice is a natural evolution of our previous research and it is our
intention to conduct further tests with people with cerebral palsy. 

Cerebral palsy has been the main concern of our research. Our main objective
is to improve their quality of life, and to help  them to be
integrated in regular schools and learning environments. That integration can 
be more effective when these people have a good technological support to overcome 
their limitations. 

The EasyVoice system is being developed as a tool for helping people with 
disabilities and is not limited to those with cerebral palsy. 
Even people without motor disorders need to have a practical user interface 
that breaks down the barriers associated with their voice disabilities. 

\section{Future work}
\label{sec:futurework}
We have much work ahead to improve the EasyVoice system so that its utilization
can be widespread worldwide. Some of the issues that we would like to work are
listed below.

\begin{itemize}
\item Incorporate voice synthesizers for other languages.
\item Implement alternative word prediction systems.
\item Design alternative user interfaces for people with very severe motor disabilities.
\end{itemize}

So far, we have only experimented EasyVoice with a voice synthesizer adapted
for the Portuguese language. We would like to extend our system to include 
voice synthesizers for other major languages, especially English, the most widely
used language in the world.

The efficiency of our word prediction can be improved as described in 
section~\ref{sec:easyvoice-tts}.

Finally, there are people with motor impairments which are so severe, that they
cannot even press a single switch. Some people are only capable of moving their
head, and others can only move their eyes. For those people, a user interface 
like ours is of little use, and we need to investigate alternative interfaces for
those cases. Some research has been done in this area~\cite{Majaranta:2002}.

\section{Summary and conclusions}
\label{sec:conclusions}
This paper presented EasyVoice, a system that combines existing technologies in
a novel way for helping people with disabilities. The software is available
for download at \url{http://w3.ualg.pt/~pcondado/download.php}.

The development of EasyVoice can be seen as a constructionist learning
experience. Just as children use LEGO blocks to create new toys, we have joined
technologies to create a new tool with a new functionality. 

Sometimes, the technology exists and is simply waiting for someone to have new 
ideas for its utilization. Most of the time, an innovation emerges from
the combination of ideas. A good example of that is the creation of the 
World Wide Web by Tim Berners-Lee. The Web came upon existence by joining 
two technologies that already existed for
quite some time: (1) the TCP/IP protocol suite, and (2) the notion of Hypertext.
The development of TCP/IP, on top of which the Web rides, already happened in
the 1970s, providing a general communications infrastructure for linking 
computers together worldwide. Likewise, the notion of Hypertext, in
which the reader is not constrained to read in any particular order, but instead
can follow links that point directly to other parts of the document, has also 
been developed in the 1960s. What Berners-Lee did, according to his own words, 
was to marry the two notions together~\cite{Berners-Lee:1999}.

Our invention is nothing compared with the creation of the Web, but it has in
itself many things in common with its creation. 
User interfaces for people with disabilities, speech synthesizers,
and VoIP applications, are all technologies that already existed for quite some time.
What we did was to marry them together, and by doing that, we believe we have created
an innovation, something that did not exist before and that opens a window for a 
new world of communications, learning, and socialization, for people with voice
and motor disabilities.


\section*{Acknowledgments}
This work was sponsored by the Portuguese Foundation for Science and Technology 
(FCT/MCES) under grant \url{POCI/CED/62497/2004}. Paulo Condado's work was also 
sponsored by Funda\c{c}\~{a}o Caloust Gulbenkian under grant Proc.~65538.

\bibliographystyle{plain}
\bibliography{references2}

\end{spacing}
\end{document}